%% file: sample-sigconf.tex
\documentclass[sigconf]{acmart}

\renewcommand\footnotetextcopyrightpermission[1]{} %

\usepackage{kotex}
\usepackage{amsthm}
\usepackage{amsmath}
\usepackage{adjustbox}
\usepackage{makecell}
\usepackage{hyperref}
\usepackage{cleveref}

\crefformat{section}{\S#2#1#3} %

\usepackage{comment}
\usepackage{booktabs}
\usepackage{diagbox}
\usepackage[english]{babel}
\usepackage[utf8]{inputenc}
\usepackage[ruled]{algorithm2e} %

\usepackage[noend]{algpseudocode}
\usepackage{algpseudocode}
\usepackage{tabularx}
\usepackage{balance}
\usepackage{array}
\newcommand{\PreserveBackslash}[1]{\let\temp=\\#1\let\\=\temp}
\newcolumntype{C}[1]{>{\PreserveBackslash\centering}p{#1}}
\newcolumntype{R}[1]{>{\PreserveBackslash\raggedleft}p{#1}}
\newcolumntype{L}[1]{>{\PreserveBackslash\raggedright}p{#1}}

\algnewcommand\algorithmicforeach{\textbf{for each}}
\algnewcommand{\algorithmicvariables}{\textbf{variables}}
\algdef{S}[FOR]{ForEach}[1]{\algorithmicforeach\ #1\ \algorithmicdo}
\algdef{SE}[VARIABLES]{Variables}{EndVariables}
   {\algorithmicvariables}
   {\algorithmicend\ \algorithmicvariables}

\theoremstyle{definition}

\newtheorem{theorem}{Theorem}%
\newtheorem{lemma}{Lemma} %

\newcommand*{\rom}[1]{\expandafter\@slowromancap\romannumeral #1@}

\AtBeginDocument{%
  \providecommand\BibTeX{{%
    \normalfont B\kern-0.5em{\scshape i\kern-0.25em b}\kern-0.8em\TeX}}}

\begin{document}

\title{Block Double-Submission Attack: Block Withholding Can Be Self-Destructive}

\author{Suhyeon Lee}
\affiliation{%
  \institution{School of Cybersecurity \\ Korea University}
  \streetaddress{Anam-ro 145}
  \city{Seoul}
  \country{Korea}}
\email{orion-alpha@korea.ac.kr}

\author{Donghwan Lee}
\affiliation{%
  \institution{Cyber/Network Technology Center \\ Agency for Defense Development}
  \city{Seoul}
  \country{Korea}}
\email{dlee@add.re.kr}

\author{Seungjoo Kim}
\affiliation{%
  \institution{School of Cybersecurity \\ Korea University}
  \streetaddress{Anam-ro 145}
  \city{Seoul}
  \country{Korea}}
\email{skim71@korea.ac.kr}

\renewcommand{\shortauthors}{Lee et al.}

\begin{abstract}

Proof-of-Work (PoW) is a Sybil control mechanism adopted in blockchain-based cryptocurrencies. It prevents the attempt of malicious actors to manipulate distributed ledgers. Bitcoin has successfully suppressed double-spending by accepting the longest PoW chain. Nevertheless, PoW encountered several major security issues surrounding mining competition. One of them is a Block WithHolding (BWH) attack that can exploit a widespread and cooperative environment called a mining pool. This attack takes advantage of untrustworthy relationships between mining pools and participating agents. Moreover, detecting or responding to attacks is challenging due to the nature of mining pools. In this paper, however, we suggest that BWH attacks also have a comparable trust problem. Because a BWH attacker cannot have complete control over BWH agents, they can betray the belonging mining pool and seek further benefits by trading with victims. We prove that this betrayal is not only valid in all attack parameters but also provides double benefits; finally, it is the best strategy for BWH agents. Furthermore, our study implies that BWH attacks may encounter self-destruction of their own revenue, contrary to their intention.
    
\end{abstract}

\ccsdesc[500]{Security and privacy~Distributed systems security}
\ccsdesc[500]{Theory of computation~Algorithmic game theory and mechanism design}

\keywords{blockchain, block double-submission attack, block withholding attack, game theoretic analysis, proof-of-work}

\maketitle

\pagestyle{plain} %

\input{section/1.introduction}

\input{section/2.related_works}

\input{section/3.system_model}

\input{section/4.analysis}

\input{section/5.gametheory}

\input{section/6.quantitative}

\input{section/7.simulation}

\input{section/8.discussion}

\input{section/10.conclusion}

\section*{Acknowledgment}
This work was supported by Institute of Information \& communications Technology Planning \& Evaluation (IITP) grant funded by the Korea government (MSIT) (No.2018-0-00532, Development of High-Assurance ($\geq$EAL6) Secure Microkernel)

\balance

\bibliographystyle{acm}
\bibliography{references}

\appendix
\section{Limit of the BWH infiltration ratio}
\label{appendix}

\begin{figure}[htb]
\includegraphics[width=0.9\linewidth]{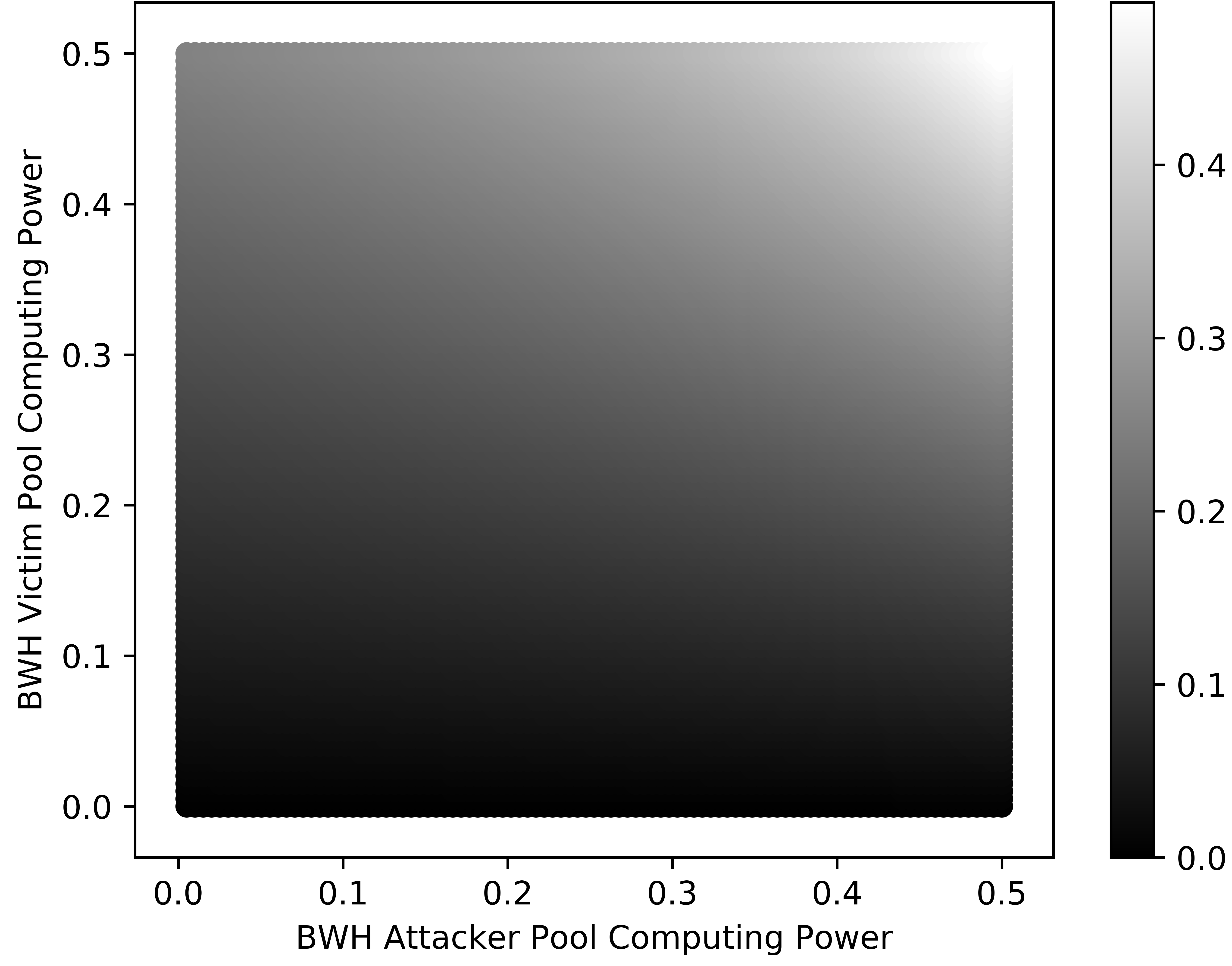}
\caption{Optimal Infiltration Ratio in the BWH attack}
\label{figure: optimal tau}
\centering
\end{figure}

As seen in Figure \ref{figure: optimal tau}, the optimal infiltration ratio $\tau$ increases as the computing power of both the attacker and victim increases. 

\begin{equation} \label{equation: optimal tau}
        f(\alpha, \beta) = \frac{\beta-\alpha\beta-\sqrt{\beta^2-\alpha\beta^2-\alpha\beta^3}}{-\alpha+\alpha^2+\alpha\beta}    .
\end{equation}

\noindent Let $\tau$ be $f(\alpha, \beta)$ defined Equation \ref{equation: optimal tau}. The value is not defined with $\alpha = \beta = 0.5$; nevertheless, we can obtain the upper limit of $\tau$ when $\displaystyle{\lim_{ \substack{\alpha \to 5-0 \\ \beta \to 5-0}} f(\alpha, \beta)}$.

\noindent For simplicity, by replacing $\alpha$ and $\beta$ by $ \displaystyle{\lim_{\delta \to +0}} (0.5 - \delta)$, we attain the following relationship:

\begin{equation}
     \displaystyle{\lim_{\alpha, \beta \to 0.5 - 0}} f(\alpha, \beta) = \displaystyle{\lim_{\delta \to +0}} \frac{-\frac{1}{2} - \delta + \sqrt{(\frac{1}{2} + 2\delta)^2 - 5\delta^2}}{2\delta}
\end{equation}

\begin{equation}
     < \frac{-\frac{1}{2} - \delta + \sqrt{(\frac{1}{2} + 2\delta)^2}}{2\delta} = \frac{-\frac{1}{2} - \delta + \frac{1}{2} + 2\delta}{2\delta} = \frac{1}{2} .
\end{equation}

\noindent Therefore, the infiltration $\tau$ is less than 0.5 when $0 < \alpha < 0.5$ and $0 < \beta < 0.5$.

\end{document}

%% file: section/1.introduction.tex
\section{Introduction} \label{section: introduction}

The effort to create true electronic cash, which began with David Chaum \cite{chaum1983blind}, has borne fruit with the emergence of blockchain. The blockchain technology enabled decentralized cryptocurrency systems, attracting several attempts to exploit the systems for illicit gain due to its nature --- the anonymity and very low participation barrier of cryptocurrency. One of the most fatal threats against a cryptocurrency system is double-spending, which spends the same digital asset more than once. To put it differently, one transaction effectively cancels out an earlier transaction, allowing the attacker to use their money twice. In such a case, the value of the cryptocurrency may plummet as consumers lose trust in the system's transactions.

The consensus of multiple entities controls a single blockchain system. As such, Sybil attacks can play a critical role in carrying out a double-spending attack. A Sybil attack occurs when an attacker gains control of many identities in a network to exert undue influence \cite{douceur2002sybil}.
There are now a variety of control techniques to mitigate Sybil attacks.
The Proof-of-Work (PoW) algorithm used in Bitcoin is the key protection against such attacks. PoW, or Proof-of-Effort, was initially developed to combat spam e-mails \cite{dwork1992pricing}. Because e-mail senders must spend sufficient time for hashing computation, PoW forces spammers to consume energy. 

On the Bitcoin blockchain, block generators, i.e., miners, are required to propose a block with a cryptographic solution on the blockchain, which requires miners to burn a significant amount of energy. A miner receives a specific quantity of Bitcoin as compensation. 
By design, the difficulty of PoW gradually increases as time goes on and more miners participate.
As a result, generating even a single block is nearly impossible for an individual miner. This barrier to entry enables Bitcoin to maintain trust in the face of a double-spending attack.

While PoW was effective at preventing a Sybil attack, several research studies have revealed that potential attacks could harm the incentive compatibility of a PoW-based blockchain system (§2).
Particularly, the selfish mining attack allowed the attacker to gain greater rewards than their mining power by purposely dissimulating a PoW block.
Furthermore, there are various attack techniques between mining pools or miners that consider real-world mining conditions, in which a few leaders established dominant possession as mining difficulty increased \cite{rosenfeld2011analysis, eyal2015miner, eyal2014majority, sapirshtein2016optimal, nayak2016stubborn, mirkin2020bdos, kwon2017selfish}.

An attacker can expect additional rewards even in a cooperative environment, mainly mining pools.
For example, Block WithHolding (BWH) attacks are effective strategies. An attacker infiltrates to another mining pool for a better profit. 
Using the victim's trust, BWH attackers only submit partial solutions (pPoW), not full solutions (fPoW), that directly reward victims --- establishing mutual trust within a mining pool is a critical issue.
There are a few methods methods to avoid such attacks \cite{eyal2014disincentivize, lee2019counter, solat2017brief, bag2017bitcoin, rosenfeld2011analysis, luu2015power, sompolinsky2015secure}; however, they cannot be fundamental remedies due to the restricted visibility of individual miners. Due to the same reason, the trust issue exists within the attacking pool. 

In this study, we propose a novel attack technique called Block Double Submission (BDS) (§3).
Under BWH attacks, a PoW block discovered by an infiltrating agent is withheld. We found that the victim pool needs the block. If the infiltrated agent, who is not supposed to withhold blocks, sells the block to the victim pool, both parties will benefit from this trade: the infiltrated agent and the victim pool. As illustrated in Figure \ref{fig: BDSA}, a BDS attack can happen when two mining pools are engaged in a BWH attack. While the infiltrated miner delivers fPoWs only to the BWH-attacking pool, the miner sells fPoWs simultaneously to two mining pools --- hence the name. 

The pivotal factor for the attack is the lack of trust between the BWH-attacking pool and its infiltrating miners, which it exploits. More generally, this attack relies on lack of trust between a mining pool and its miners like the BWH attack does.
Compared to the benefits from a BWH attack, which are limited to 10\%, the rewards from a BDS attack can reach nearly 100\%. This double reward strongly motivates the BDS attackers. 
The victim pool can recover its losses from the BWH attack through this new trade by publishing fPoWs, and the traitor agent can make a substantial profit.

As previously stated, this new strategy damages Block WithHolding (BWH)-attacking pools. Numerous research has explored a BWH-based counterattack to the BWH attack \cite{bag2017bitcoin, lee2019counter, luu2017smartpool, rosenfeld2011analysis}. Detecting and assessing the infiltration power of a BWH attack is a critical prerequisite for conducting the counterattack. However, detecting a BWH attack from a private pool is nearly impossible \cite{lee2019counter}. Big private mining pools exist in the real world, for example, BitFury. We cannot join this pool freely and cannot get proof of attacks even if an attack is underway. On the other hand, betrayal from inside is possible even in a closed group. In other words, traitors in a BWH-attacking pool can ruin the pool's strategy for their additional benefit.

To demonstrate that such a betrayal deal is viable, we establish three BDS attack conditions and prove that trades satisfying the criteria always exist (§4). Then, using game-theoretic approaches, we show that mining agents' dominant strategy is to betray the BWH attacking pool and participate in a BDS attack (§5). Consequently, we demonstrate that a BWH attacker suffers a loss due to the principal-agent problem between a BWH attacking pool and mining agents. 
The quantitative study (§6) and simulations (§7) indicate the BDS attack's impact and corroborate our theoretical analysis. BDS is a subtle attack scheme that damages another attacker while compensating for the victim's loss. Finally, we examine BDS's multilateral aspects and applications in discussion (§8).

This paper presents the BDS attack, raising the question of whether a mining pool can trust its miners unconditionally. Our contributions are as follows:

\begin{itemize}
    \item We propose a novel attack technique, the BDS attack, in which a mining agent betrays the BWH attacking pool in pursuit of more rewards.
    \item We demonstrated that a BDS attack could reward the betrayed agent with twice as much benefit as merely participating in a BWH attack.
    \item We discovered that a BWH attack could result in significant losses to a BWH attacking pool due to BDS attacks.
\end{itemize}

%% file: section/2.related_works.tex
\section{Related Work} \label{section: related work}

This section reviews two related attacks, selfish mining and BWH attacks, on PoW mining.

\subsection{Selfish Mining}

Selfish mining is the process of retaining blocks to force other miners to squander computational power. In brief, the attacker discovers $N (N > 2)$ blocks first and then publishes them when other miners discover $K (N > K)$ blocks. Due to the long chain rule, the blocks broadcasted by other miners do not belong to the main chain. 
Therefore, the attacker can publish blocks above their proportional mining power.
As a block rewards the miner who published it, the attacker can make a disproportionate profit.

Eyal and Sirer initially proposed the concept of selfish mining \cite{eyal2014majority}. This study shows that selfish mining can generate disproportionate revenue for attackers. 
Sapirshtein et al. \cite{sapirshtein2016optimal} found the optimal conditions for selfish mining. 
It demonstrated that selfish miners could generate more revenue by employing dynamic methods.
An eclipse attack isolates a particular peer-to-peer network node. 
Nayak et al. \cite{nayak2016stubborn} suggested a mining strategy incorporating an eclipse attack and selfish mining to increase revenue. 
Ethereum used a modified version of the Greedy Heaviest Object Subtree (GHOST) protocol and rewards of uncle blocks for its safe, high-throughput environment \cite{sompolinsky2015secure}. Liu and Feng demonstrated that Ethereum is similarly susceptible to selfish mining.

There are studies on the countermeasures against selfish mining \cite{solat2016zeroblock}, \cite{pass2017fruitchains}. They are primarily associated with negating or delayed-publication blocks. Lee and Kim \cite{lee2022rethinking} suggested a selfish mining detection and mitigation technique in a mining pool environment. 
Like the BWH attack detailed in the following section, selfish mining involves the withholding of blocks. Also, as the BWH attack does not require much CPU power, it may be more viable in a real-world mining setting.

\subsection{Block Withholding Attack}

A BWH attack is a strategy that miners delay the submission of blocks in PoW mining environments, generalizing selfish mining. Specifically, a BWH-attacking pool delays the submission of blocks in the target mining pool. Our research focuses solely on the latter concept, namely mining pool behavior. 

The concept of a mining pool is to enable collaboration, as it is exceedingly unusual for individuals to discover blocks. Mining pools facilitate PoW tasks for miners. 
Miners discovering nonces for a hash value in the mining pool are compensated for them, even if the nonces do not satisfy the networks' difficulty requirements.
This is known as a pPoW, and a valid nonce in the network is referred as fPoW. Mining pools get rewards from the PoW blockchain network via fPoW and distribute them to their miners.

Rosenfeld \cite{rosenfeld2011analysis} conceived the initial idea for a BWH attack. 
The study presented an attack concept sabotaging the mining pool by submitting only pPoWs, unrelated to miners receiving rewards from the Bitcoin network.
Eyal \cite{eyal2015miner} proposed an attack strategy advantageous to the attacker but detrimental to the target. This attack strategy comprises mining pools infiltrating other mining pools and some miners submitting only pPoWs to indirectly diminish the effectiveness of other mining pools. When two mining pools engage in attacks, each suffers a loss. Kwon et al. \cite{kwon2017selfish} addressed this dilemma by employing the Fork After Withholding (FAW) technique. Their work introduced advantageous attack technique since it only submits blocks under forking situations rather than always withholding them. In addition, Liu et al. \cite{liu2020evaluation} and Chang et al. \cite{chang2019uncle} examined an attack technique leveraging an uncle block rewarded in Ethereum.

Countermeasures against the BWH attack can be classified primarily into those that modify mining mechanisms and others. In order to prevent miners from distinguishing between fPoW and pPoW, Bag et al. \cite{bag2017bitcoin} and Eyal \cite{eyal2014disincentivize} recommended cryptographically splitting the mining mechanism into two stages. Lee and Kim \cite{lee2019counter} developed a strategy for detecting and responding to the BWH attack by installing sensor miners in different mining pools. Sarker et al. \cite{sarker2019anti} presented a mining pool-wide anti-withholding reward system to minimize BWH attacks' efficacy by increasing honest miners' rewards.

Our research uses a significantly different strategy than the previous studies. The study provides both an offensive and a defensive technique aimed towards attackers. In addition, our solution is realistic because it does not require alterations to the fundamental PoW mining procedures. The following section details a system model and problem statements.

%% file: section/3.system_model.tex
\section{System Model and Problem Statement}    \label{section: system model and problem}

    This section illustrates the system model under consideration by explaining Bitcoin PoW miners and mining pools. Then we define the problem statement that raises a trust concern about the BWH attack.

    \subsection{System Model}
    
    The analysis presented in this paper assumes that at least two mining pools exit on the Bitcoin network. Miners and mining pools can easily enter and exit a mining pool. Algorithm \ref{algorithm: mining in a pool} depicts the algorithm for mining in a mining pool. A mining pool assigns a miner a task, including an information set block to mine. 
    The miner attempts to solve the cryptographic puzzle using hash functions and nonce values.
    The miner can submit two types of PoW, including $pPoW$ and $fPoW$. Miners receive rewards proportional to their PoW contribution from the mining pool.

    \begin{algorithm}[tb]
    \caption{Mining in a Mining Pool $\mathcal{A}$}\label{algorithm: mining in a pool}
    \SetKwProg{Mining}{Function \emph{Mining}}{}{end}
    \Mining{}{
        task $\gets$ \textbf{\textsf{newTask}}($w$)\;
        (pPoW, fPoW) $\gets$ \textbf{\textsf{work}}(task)\;
        \textbf{\textsf{send}}($\mathcal{A}$, (pPoW, fPoW))\;
        revenue $\gets$ revenue + \textbf{\textsf{recv}}( $\mathcal{A}$ )\;
    }
    \end{algorithm}
    
    Algorithm \ref{algorithm: mining pool reward} outlines the mining pool rewards mechanism. A mining operation offers miners $fPoW$ and $pPoW$, and then it publishes $fPoW$ as a legitimate block to the Bitcoin network. The block contains the coinbase transaction of the mining pool, which mints new Bitcoins for the mining pool. The payout for miners is proportional to the number of $pPoW$. In other words, submitting $fPoW$ does not result in a greater reward than submitting $pPoW$.
    
    \begin{algorithm}[tb]
    \caption{Mining Reward in a Mining Pool}\label{algorithm: mining pool reward}
    \SetKwProg{Reward}{Function \emph{Reward}}{}{end}
    \Reward{}{
        \ForAll{Miner $w$ $\in$ \text{Miners}}{
            (pPoW, fPoW) $\gets$ \textbf{\textsf{recv}}($w$)\;
            revenue $\gets$ revenue + \textbf{\textsf{publish}}(fPoW)\;
            reward($a$) $\gets$ \textbf{\textsf{count}}(pPoW)\;
        }
    }
    \end{algorithm}

    In the BWH attack, a mining pool infiltrates another pool of miners. The former is known as a BWH-attacking pool, whereas the latter is known as a victim mining pool or victim pool. In a BWH-attacking pool, miners performing BWH mining are called BWH miners. Infiltrated miners do not contribute fPoW to the victim pool. Consequently, both the victim pool's earnings and the infiltrating miners' incentive from the victim pool decrease. As the total mining power diminishes; however, the BWH-attacking pool's revenue increases more than the loss in infiltrating miners. 
    
    Given that the total network mining power is 1, let the BWH-attacking pool's computing power be $\alpha$ and the victim pool's computing power be $\beta$. They must not exceed 50\%, i.e., $0 \leq \alpha, \beta \leq 0.5$. Let the infiltration ratio from the attacker to the victim pool be $\tau$ where $0 \leq \tau \leq 1$. Then the revenue of the BWH-attacking pool is given as follows:

    \begin{equation}    \label{equation: BWH-attacking pool before}
        \frac{(1-\tau)\alpha}{1-\tau\alpha} + \frac{\beta}{1-\tau\alpha} \frac{\tau\alpha}{\beta + \tau\alpha} .
    \end{equation}

    \noindent The revenue of the victim pool is as follows:

    \begin{equation} \label{equation: victiim pool before}
        \frac{\beta}{1-\tau\alpha} \frac{\beta}{\beta + \tau\alpha}.
    \end{equation}

    \noindent Then, we consider a BWH attack performed with the optimal infiltration ratio $\tau$ \cite{lee2019counter}, which is:

    \begin{equation}
        \frac{\beta-\alpha\beta-\sqrt{\beta^2-\alpha\beta^2-\alpha\beta^3}}{-\alpha+\alpha^2+\alpha\beta}.
    \end{equation}
    
\begin{figure*}[htb]
    \caption{Structure of the naïve BWH attack: The miners infiltrate Pool \textbf{B} (the victim pool), following the order of Pool \textbf{A} (the BWH attacker). The BWH-attacking pool cannot observe nor can expect trustworthy BWH behaviors from the infiltrated miners.}
    \label{fig: Naive BWH}
    \centering
    \includegraphics[width=0.7\linewidth]{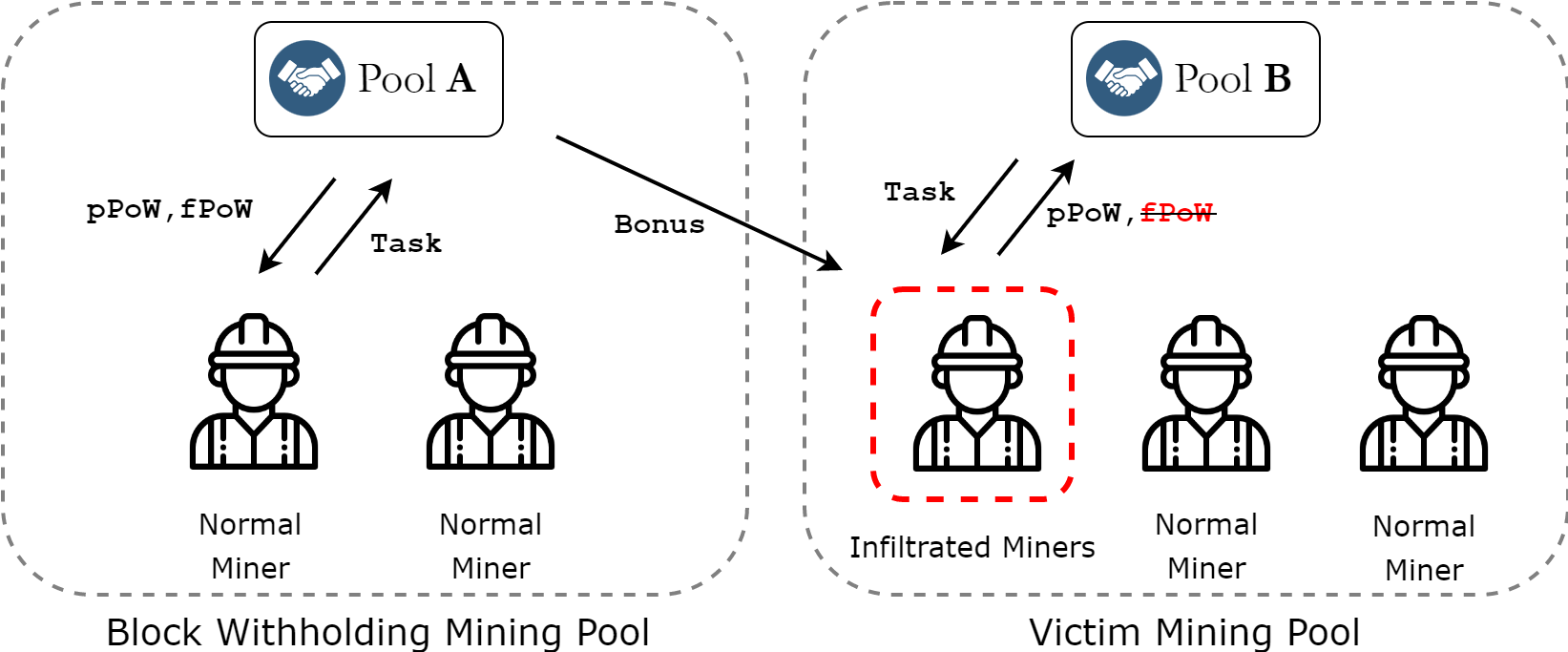}

    \caption{Structure of the fixed BWH attack: The miners infiltrate Pool \textbf{B} (the victim pool) through Pool \textbf{A} (the BWH attacker), where the BWH-attacking pool can control the BWH behaviors of the infiltrated miners and fPoWs are not submitted to Pool \textbf{B}. The infiltrated miners may betray the attacking pool, and Pool \textbf{B} requires the withheld fPoWs.}
    \label{fig: fixed BWH}
    \centering
    \includegraphics[width=0.7\linewidth]{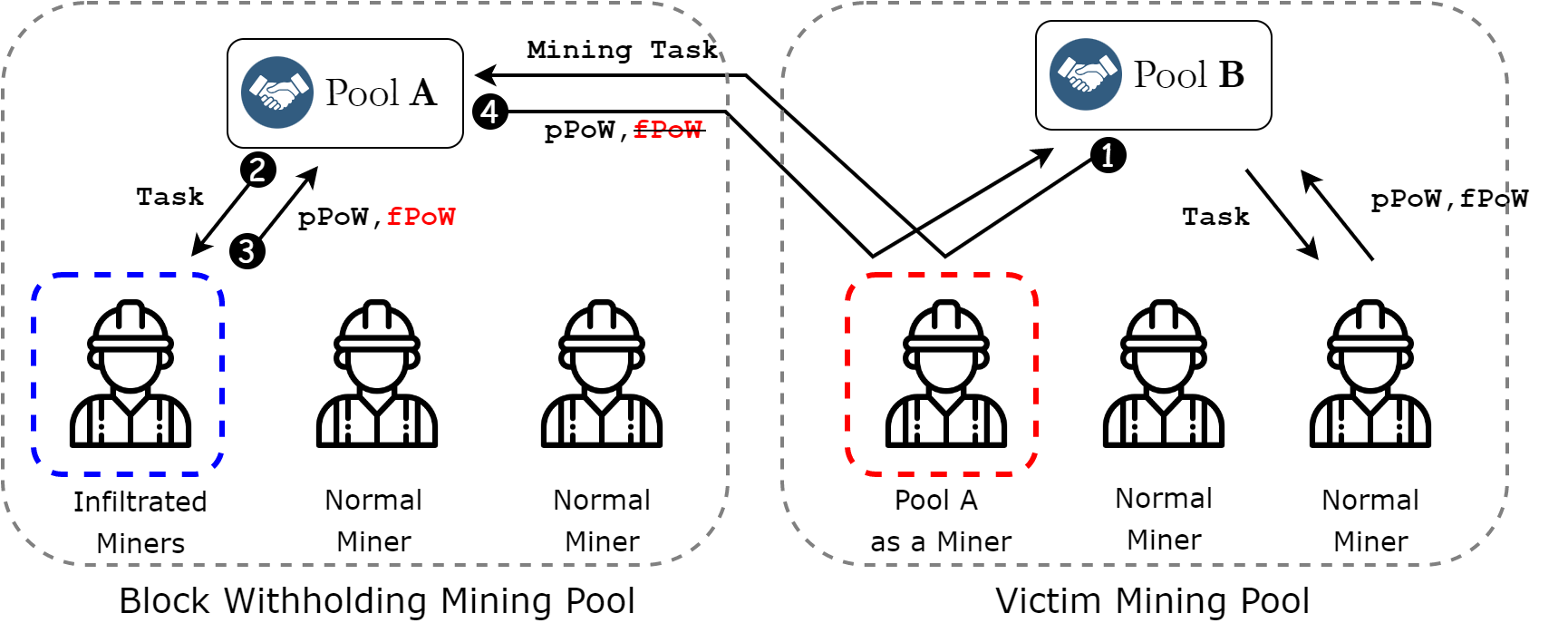}

    \caption{BDS Attack: The miners infiltrate Pool \textbf{B} (victim), following the order from Pool \textbf{A} (the BWH attacker). The infiltrated miners submit PoWs to Pool \textbf{A} and Pool \textbf{B} simultaneously, and Pool \textbf{B} can publish fPoWs, which might also be withheld, making an extra profit. The infiltrated miners can share the extra profit by selling fPoWs to the victim pool.}
    \label{fig: BDSA}
    \centering
    \includegraphics[width=0.7\linewidth]{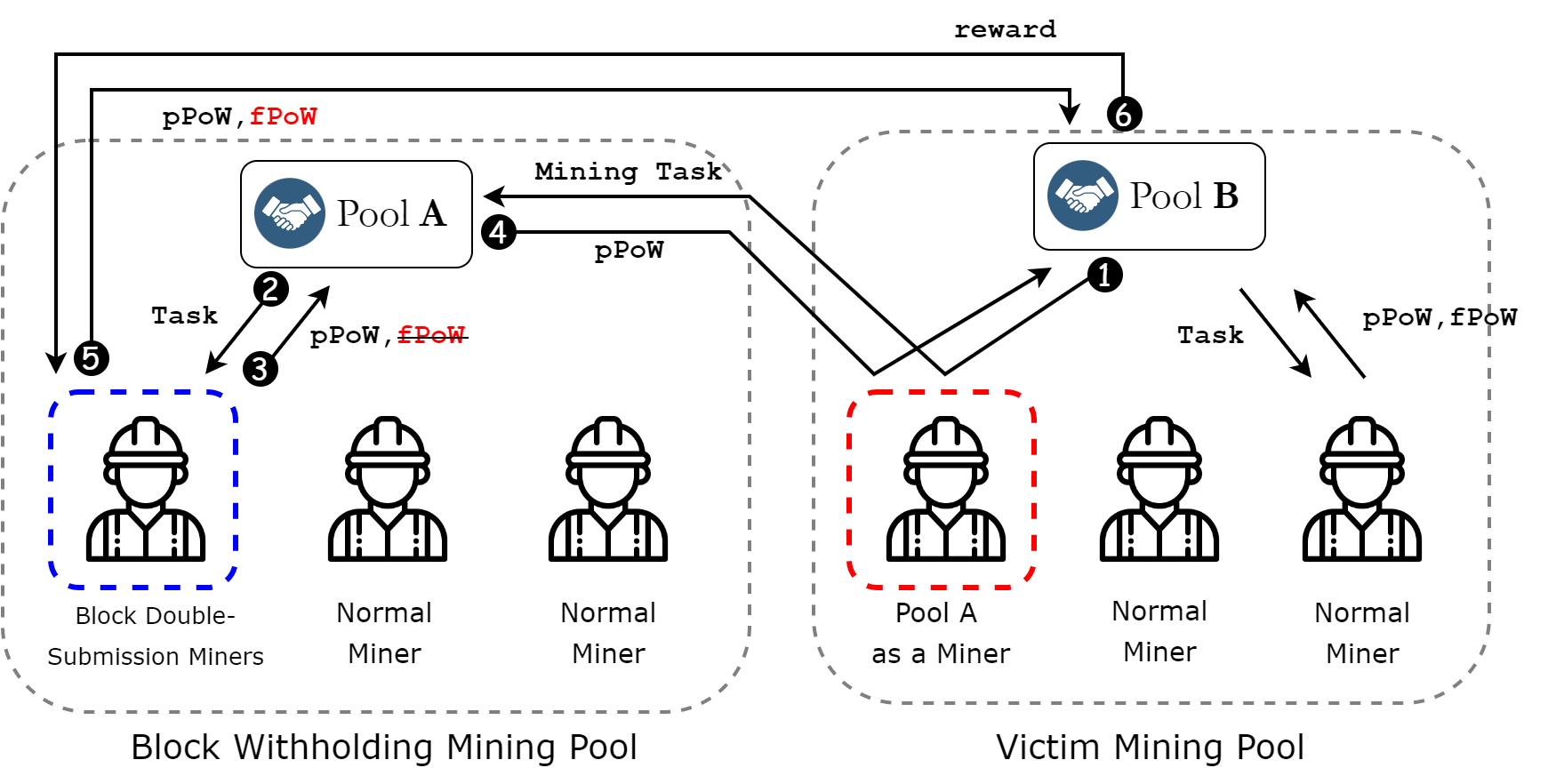}
    \end{figure*}

    From a structural viewpoint, a BWH-attacking pool can remotely infiltrate BWH miners into a victim pool and share earnings. Figure \ref{fig: Naive BWH} illustrates this concept, which is referred to as a naive block withholding structure. Pool \textbf{A} is the attacking BWH pool, whereas Pool \textbf{B} is the victim BWH pool. The infiltrated miners who participated do not submit fPoW to sabotage Pool \textbf{B}'s revenue. Instead, Pool \textbf{A} divides its increased earnings with the infiltrated miners to compensate for Pool \textbf{B}'s diminished rewards.

    The best response for a victim mining pool is a mutual BWH attack against the attacking pool, damaging both mining pools.
    On the other hand, detecting a BWH attack in the real world might be challenging. Research about detection \cite{lee2019counter} pointed out that it is impossible to detect a BWH attack using a private mining power with absolute certainty. In this vein, we can presume that this difficulty in detecting the BWH attack comes from the incapability of launching an effective counterattack. This study, therefore, assumes that it is practically hard for the victim pool to counterattack the BWH-attacking pool.

\subsection{Problem Statement}
    
    The BWH attack exploits the lack of trust between a mining pool and its users. The attack can succeed if the infiltrating miners send only pPoW to the BWH-attacking pool in cooperation. We suspect, however, that the level of trust between the BWH-attacking pool and BWH miners is sufficient to lead such cooperation.

    The BWH-attacking pool cannot fully regulate the operations of infiltrated miners, given the current structure of mining pools. For instance, by submitting fPoW to the victim pool and disobeying the direction from the attacking pool, the infiltrated miners can still receive rewards from the BWH-attacking pool. With this behavior, the infiltrating miners likely generate more revenue than the intended BWH attack. Consequently, the BWH mining attack against the pool will be in vain.

    Nevertheless, there is still a significant problem. Even though the BWH-attacking pool directly penetrates the victim pool, internal miners can still trade fPoWs with the victim pool. A mining task contains coinbase information, which indicates which mining pool is currently under attack. Based on this information, miners can submit fPoW independently of the BWH-attacking pool they belong to. Existing mining pools' incentive schemes typically do not differentiate between fPoW and pPoW. In this case, however, additional awards for fPoWs may be necessary because they provides an incentive to deviate from their ordinary behavior. 
    It would also allow BWH miners who merely benefited from utilizing the reduced mining power via sabotage to make significantly more rewards than BWH attack-only cases.

    Figure \ref{fig: BDSA} illustrates this strategy.
    We refer to the strategy as a \textbf{Block Double-Submission (BDS) Attack} in which PoW is submitted to both a BWH-attacking pool and a victim pool. 
    BDS miners receive concurrent payments from both mining pools through such a method.
    Specifically, the BDS miners submit only pPoW to the BWH-attacking pool, provided they have successfully identified the victim pool. If the victim pool publishes a fPoW previously submitted to the BWH-attacking pool, the BWH-attacking pool could identify the traitor. The victim pool receives pPoW and fPoW from BDS miners but does not require pPoW to mint coins. In this instance, pPoWs measure the mining power of traitors. In the following section, we examine the rationality of the BDS attack.

%% file: section/4.analysis.tex
\section{Block Double-Submission Attack}  \label{section: rationality analysis}

    This section will explain the rationale behind the BDS attack.
    First, we identify the conditions that must be met for the attack to be reasonable, and then we prove that they are feasible. 
    Second, we present algorithms for the BDS attack.

    \subsection{Rationality of the Block Double-Submission}

For this attack to be reasonable, it must be mutually advantageous and have a minimal boundary.
The block double-submission attack must provide BDS miners greater benefits (\textbf{C1)}. Similarly, it ought to be profitable for the victim pool (\textbf{C2}). In addition, we require that the victim pool cannot provide higher rewards than a BWH miner receives for mining with integrity (\textbf{C3}). Accordingly, we summarize three requirements as follows:

\begin{itemize}
    \item \textbf{(C1)} The trade generates greater profits for the victim pool than the honest mining under a BWH attack.
    \item \textbf{(C2)} The trade provides the BWH miner greater revenue than cooperation for a BWH attack.
    \item \textbf{(C3)} The trade is established with rewards less than the ones given to the BWH miner by honest mining.
\end{itemize}

\noindent Let a price function for purchasing a valid block from a BWH miner with mining power $p$ where $p \leq \tau\alpha$ be $\mathcal{T}(p)$, and the revenue generated by a victim pool is:

\begin{equation} \label{equation: victim pool after}
    \biggl[ \overbrace{ \frac{\beta + p}{1-\tau\alpha+p} }^{p \text{ from a BWH miner}}
    - \overbrace{ \mathcal{T}(p) }^{ \text{cost for the trade} } \biggr]
    \cdot \overbrace{ \frac{\beta}{\beta + \tau\alpha} }^{ \text{The BWH-attacking pool is still infiltrating} } .
\end{equation}

\noindent Then, the revenue of a BWH miner with the trade is calculated as follows:

\begin{multline} \label{equation: BWH miner after}
    \overbrace{ \mathcal{T}(p) }^{ \text{revenue from the trade} }
    + \frac{p}{\alpha} \cdot \biggl[ 
    \overbrace{ \frac{(1-\tau)\alpha}{1-\tau \alpha + p} }^{ \text{revenue from the BWH-attacking pool} } \\
    + \overbrace{ \Big\{\frac{\beta + p}{1-\tau \alpha + p} - \mathcal{T}(p) \Big\}  \frac{\tau \alpha}{\beta + \tau \alpha} }^{ \text{revenue from the victim pool} \text{   } }
    \biggr]   .
\end{multline}

\noindent Using the above, we obtain the boundaries of $\mathcal{T}(p)$ that satisfies \textbf{C1}, \textbf{C2}, and \textbf{C3}, respectively.

\begin{lemma} \label{lemma: C1}
    There exists a boundary of $\mathcal{T}(p)$ for \textbf{C1}.
\end{lemma}

\begin{proof}
    
    For \textbf{C1}, $\mathcal{T}(p)$ must satisfy $R_{t, v} > R_{b, v}$, where $R_{t, v}$ is the revenue with the trade and $ R_{b, v}$ is the revenue without the trade. $R_{t, v}$ is given in (\ref{equation: victim pool after}). $ R_{b, v}$ is provided as (\ref{equation: victiim pool before}) in Section \ref{section: system model and problem}. Therefore, the following should hold:
    
    \begin{equation}
        \frac{\beta}{1-\tau\alpha} \cdot \frac{\beta}{\beta+\tau\alpha} < \biggl[ \frac{\beta + p}{1-\tau\alpha+p} - \mathcal{T}(p) \biggr] \cdot \frac{\beta}{\beta + \tau\alpha}    .
    \end{equation}
    
    \begin{equation}
        \mathcal{T}(p) < \frac{p\cdot(1-\tau\alpha-\beta)}{(1-\tau\alpha+p)(1-\tau\alpha)}
    \end{equation}
\end{proof}

\begin{lemma} \label{lemma: C2}

There exists a boundary of $\mathcal{T}(p)$ for \textbf{C2}.

\end{lemma}
\begin{proof}

For \textbf{C2}, $\mathcal{T}(p)$ must satisfy $R_{t, m} > R_{b, m}$, where $R_{t, m}$ is the revenue with the trade and $ R_{b, m}$ is the revenue without the trade.

$R_{t, m}$ was shown previously in (\ref{equation: BWH miner after}). $ R_{b, m}$ is given as follows:

\begin{equation}
    R_{b, m} = \frac{p}{\alpha} \cdot \biggl[ \frac{(1-\tau)\alpha}{1-\tau \alpha} + \frac{\beta}{1-\tau \alpha} \frac{\tau \alpha}{\beta + \tau \alpha} \biggr]   .
\end{equation}

\noindent Therefore, the following must hold:

\begin{multline}
    \mathcal{T}(p) + 
    \frac{p}{\alpha} \cdot \biggl[ \frac{(1-\tau)\alpha}{1-\tau \alpha + p} + \\
    \Big\{\frac{\beta + p}{1-\tau \alpha + p} - \mathcal{T}(p) \Big\}  \frac{\tau \alpha}{\beta + \tau \alpha} \biggr]  > \\
    \frac{p}{\alpha} \cdot \biggl[ \frac{(1-\tau)\alpha}{1-\tau \alpha} + \frac{\beta}{1-\tau \alpha} \frac{\tau \alpha}{\beta + \tau \alpha} \biggr] .
\end{multline}

\begin{multline}
    \mathcal{T}(p) > \frac{\beta+\tau\alpha}{\beta+\tau\alpha-p\tau} \frac{1}{(1-\tau\alpha)(1-\tau\alpha + p)} \frac{p}{\alpha} \cdot \\
    \biggl[ p (1-\tau) \alpha + \frac{\tau\alpha}{\beta+\tau\alpha} \cdot \big( p\beta - p (1-\tau\alpha)\big) \biggr]
\end{multline}

\end{proof}

\noindent Then we prove that \textbf{C3} is satisfied comprehensively when $\mathcal{T}(p)$ is already within the boundary for \textbf{C1}.

\begin{lemma}     \label{lemma: C3}
    If $\mathcal{T}(p)$ satisfies \textbf{C1}, $\mathcal{T}(p)$ also satisfies \textbf{C3}.
\end{lemma}

\begin{proof}
    \noindent From Lemma \ref{lemma: C1}, the boundary of $\mathcal{T}(p)$ is given as follows:
    
    \begin{multline}
        \mathcal{T}(p) < \frac{p\cdot(1-\tau\alpha-\beta)}{(1-\tau\alpha+p)(1-\tau\alpha)}    \\
        = p \cdot \frac{1-\tau\alpha-\beta}{1-\tau\alpha+p-\tau\alpha+\tau^2\alpha^2 - p\tau\alpha}.
    \end{multline}
    
    \noindent By using the chain of inequality, which is:
    
    \begin{equation} \label{inequality: assumption}
        p \leq \tau\alpha < \beta < 0.5 ,
    \end{equation}

    \noindent then we attain the following:
    
    \begin{equation}
        \beta > -p + \tau\alpha + p\tau\alpha - \tau^2\alpha^2 .
    \end{equation}

    \noindent Therefore, the following relation holds:
    
    \begin{equation}
         p \cdot \frac{1-\tau\alpha-\beta}{1-\tau\alpha+p-\tau\alpha+\tau^2\alpha^2 - p\tau\alpha} < p   .
    \end{equation}
    
    \noindent As $\mathcal{T}(p) < p \leq \tau\alpha < \beta$, the maximum of $\mathcal{T}(p)$ is not greater than $\tau\alpha$.

\end{proof}

\noindent From the previous results, we can prove that all the conditions are satisfied under our system model.

\begin{theorem} \label{theorem: condition statisfaction}

\noindent There exists $\mathcal{T}(p)$ satisfying all the conditions \textbf{C1--3} at the same time.

\end{theorem}

\begin{proof}
    With Lemma \ref{lemma: C3}, we only need to prove that there exists $\mathcal{T}(p)$ that satisfies \textbf{C1} and \textbf{C2} at the same time. According to Lemma \ref{lemma: C1} and Lemma \ref{lemma: C2}, $\mathcal{T}(p)$ must satisfy the below inequality.
    
    \begin{multline}
    \frac{p\cdot(1-\tau\alpha-\beta)}{(1-\tau\alpha+p)(1-\tau\alpha)} > \mathcal{T}(p) > \\
    \frac{\beta+\tau\alpha}{\beta+\tau\alpha-p\tau} \frac{1}{(1-\tau\alpha)(1-\tau\alpha + p)} \frac{p}{\alpha} \cdot \\
    \biggl[ p (1-\tau) \alpha + \frac{\tau\alpha}{\beta+\tau\alpha} \cdot \big( p\beta - p (1-\tau\alpha)\big) \biggr]
    \end{multline}
    
    \noindent By simplifying the above equation, we obtain the following relation:
    
    \begin{multline}
        1 - \tau\alpha - \beta > \\ p \cdot \frac{\beta+\tau\alpha}{\beta+\tau\alpha-p\tau} \biggl[ (1-\tau) + \frac{\tau}{\beta+\tau\alpha} \cdot \big( \beta - (1-\tau\alpha)\big) \biggr]
    \end{multline}

    \begin{equation}
        = p + \frac{\overbrace{\tau (p - 1)}^{p -1 < 0}}{ \beta+\tau\alpha - p \tau}  .
    \end{equation}
    
    \noindent Using $\tau < 0.5$ (See Appendix \ref{appendix} for a proof), we can prove the following:
    
    \begin{multline}
        1 - \tau\alpha - \beta > \tau\alpha \geq p > p + \frac{\overbrace{\tau (p - 1)}^{p -1 < 0}}{ \beta+\tau\alpha - p \tau}
    \end{multline}
    
    \noindent Using (\ref{inequality: assumption}), we finish the proof.
    
    \begin{equation}
        1  > \alpha + \beta > 2\tau\alpha + \beta
    \end{equation}
    
\end{proof}

    Eventually, we proved the existence of a rational deal that satisfies all the given conditions. The BDS attack block is advantageous for both parties. The victim mining pool can acquire fPoW without paying the total rewards associated with its publishing. 
    As a result, the BDS attack can reasonably be contracted and executed between BWH miners and a victim pool.

    \subsection{Algorithms for Block Double-Submission Attack}
    
    We briefly explained the mechanism of the block-double submission attack in a problem statement. Now, we will discuss specific algorithms. Although the block-double submission has been proven reasonable, the current mining pool model needs to be modified for the BDS trade. The BDS attacker, for example, submits \textit{fPoW} to the BWH victim pool. According to Algorithm \ref{algorithm: mining in a pool}, fPoW does not affect a miner's rewards. Moreover, even though a mining pool counts an $fPoW$ as the rewards by a certain number of $pPoW$s, the proper calculation of the rewards boundary demands the BWH miner's mining power. The BWH victim pool will, therefore, only accept the block-double submission if $fPoW$ exists and the reward is proportional to the number of $pPoW$s.

    Algorithm \ref{algorithm: BDS mining in a BWH-attacking pool} describes the BDS miner. The BDS deal is established only when \textit{fPoW} already exists. When the miner discovers \textit{fPoW}, he sends both \textit{pPoW} and \textit{fPoW} to both the BWH-attacking pool and the victim pool. If not, only \textit{pPoW} is submitted to the BWH-attacking pool.

    Algorithm \ref{algorithm: victim pool reward} explains how the BWH victim pool computes the rewards for a BDS miner. First, the victim pool disregards the miner's rewards if \textit{fPoW} is omitted. Otherwise, the mining pool calculates the mining payout using the number of \textit{pPoW}. The number of \textit{pPoW} directly indicates the miner's computational power in a probabilistic manner.

    \begin{algorithm}[tb]
    \caption{BDS between the BWH-attacking pool \textbf{A} and the BWH victim pool \textbf{B}}
    \label{algorithm: BDS mining in a BWH-attacking pool}

    \SetKwProg{Mining}{Function \emph{Mining}}{}{end}
    \Mining{}{
        task $\gets$ \textbf{\textsf{newTask}}($w$)\;
        (pPoW, fPoW) $\gets$ \textbf{\textsf{work}}(task)\;
        \eIf{fPoW $\neq$ \textsf{null}}
        {
            $\textbf{\textsf{send}}(\textbf{A},(\text{pPoW}))$\;
            $\textbf{\textsf{send}}(\textbf{B},(\text{pPoW,fPoW}))$\;
        }{
            $\textbf{\textsf{send}}(\textbf{A},(\text{pPoW}))$\;
        }
        revenue $\gets$ revenue + \textbf{\textsf{recv}}(\textbf{A}) + \textbf{\textsf{recv}}(\textbf{B})\;
    }
    \end{algorithm}

    \begin{algorithm}[tb]
    \caption{Rewards from the BDS attack} \label{algorithm: victim pool reward}

    \SetKwProg{Reward}{Function \emph{Reward}}{}{end}
    \Reward{}{
        \ForAll{Miner $w$ $\in$ \text{Miners}}{
            (pPoW, fPoW) $\gets$ \textbf{\textsf{recv}}($w$)\;
            \If{$\text{fPoW} = \textsf{null}$}
            {
                \textbf{\textsf{quit}}()\;
            }
            
            revenue $\gets$ revenue + \textbf{\textsf{publish}}(fPoW)\;
            reward($a$) $\gets$ \textbf{\textsf{trade}}\textbf{\textsf{Count}}(pPoW)\;
        }
    }
    \end{algorithm}

%% file: section/5.gametheory.tex
\section{Game-Theoretical Analysis} 
\label{section: game theory analysis}

This section examines the double-submission block attack from a game-theoretic perspective. First, we investigate the strategies employed by miners in the BWH-attacking pool. The second step is investigating the pricing of fPoW submitted to the victim pool. Using the preceding findings, we demonstrate that a BWH attack  may be detrimental to the BWH attackers. In short, the dominant strategies of BWH miners are to execute BDS attacks for profit, resulting in the loss of the BWH-attacking pool.

\begin{table}[htb]
\centering
\caption{Payoff table of two miners in a BWH-attacking pool}
\label{table: utility table of two miners bwh betrayal game}
\adjustbox{width=0.6\linewidth}{%
\begin{tabular}{c|c|c|}
\cline{2-3}
                        &   C   &  B    \\ \hline
\multicolumn{1}{|c|}{Cooperate (C)} & $R, R'$ & $D, H'$ \\ \hline
\multicolumn{1}{|c|}{Betray (B)} & $H, D'$ & $L, L'$ \\ \hline
\end{tabular}
}
\end{table}

\subsection{A Cooperation Game between BWH Miners} \label{subsection: cooperation between miners}

The previous section showed that the BDS trade between a miner and a victim pool can be established reasonably. Given that multiple BWH miners may exist in a BWH-attacking pool, it is important to analyze the dominant strategy as the collective actions of multiple miners to better understand the miners' behaviors.

We begin by analyzing a simple model with two miners. Assuming the simplest set of BWH miners, denoted by $M = \{m_1, m_2\}$,  each miner is assigned the action set $A = \{C, B\}$ where $C$ represents `Cooperate' and $B$ represents `Betray' (i.e., the BDS). Table \ref{table: utility table of two miners bwh betrayal game} is the payoff matrix for the two BWH miners game. Specifically, we represent the mining power of $m_1$ as $p$ and $m_2$ as $q$, where $p + q \leq \tau \alpha$.

\begin{lemma} \label{lemma: inequality}
In Table \ref{table: utility table of two miners bwh betrayal game}, the boundary of $\mathcal{T}(\cdot)$ by Theorem \ref{theorem: condition statisfaction} satisfies the following chains of inequalities:
    \begin{equation}
        H > R > D \text{ and } L > R > D
    \end{equation}
    \begin{equation}
        H' > R' > D' \text{ and } L' > R' > D'
    \end{equation}
\end{lemma}

\begin{proof}

    First, for $H > R$, $\mathcal{T}(\cdot)$ by Lemma \ref{lemma: C1} and \ref{lemma: C2} satisfies it. \newline
    
    Second, for $R > D$, $R$ and $D$ are as the following:
    
    \begin{equation}
        R = \frac{p}{\alpha} \cdot \biggl[ \frac{(1-\tau)\alpha}{1-\tau \alpha} + \frac{\beta}{1-\tau \alpha} \frac{\tau \alpha}{\beta + \tau \alpha} \biggr] 
    \end{equation}
    
    \begin{equation} \label{value: D}
        D = \frac{p}{\alpha} \cdot \biggl[ \frac{(1-\tau)\alpha}{1-\tau \alpha + q} + \Big\{\frac{\beta + q}{1-\tau \alpha + q} - \mathcal{T}(q) \Big\} \frac{\tau \alpha}{\beta + \tau \alpha} \biggr] 
    \end{equation}
    
    \begin{multline}
        R - D = \frac{p}{\alpha} \cdot \biggl[ 
        \Big\{ \frac{1}{1-\tau \alpha} - \frac{1}{1-\tau \alpha + q} \Big\} 
        \cdot (1-\tau)\alpha + \\ 
        \Big\{ \frac{\beta}{1-\tau \alpha} -\frac{\beta + q}{1-\tau \alpha + q} \Big\} \cdot \frac{\tau \alpha}{\beta + \tau \alpha} \biggr]  + \mathcal{T}(q) \frac{p\tau }{\beta + \tau \alpha} \\
        = \frac{p}{\alpha} \cdot \frac{q}{(1-\tau\alpha)(1-\tau\alpha+q)} \cdot \frac{\beta}{\beta + \tau \alpha}   + \mathcal{T}(q) \frac{p\tau }{\beta + \tau \alpha} 
    \end{multline}

    $R>D$ is always satisfied with $\mathcal{T} > 0$. Therefore, by Lemma \ref{lemma: C1} and Lemma \ref{lemma: C2}, the boundary of $\mathcal{T}(\cdot)$ satisfies $H > R > D$. $H' > R' > D'$ is satisfied likewise. \newline
    
    Third, we show $L > R$. In the strategy $(B, B)$, BWH miners $m_1$ and $m_2$ betray their mining pool. For the sake of analysis, we consider it as that a miner with a collective BWH mining power of $p' = p+q$ betrays the BWH-attacking pool. By Lemma \ref{lemma: C1} and \ref{lemma: C2}, the relationship $L > R$ holds. We already showed $R > D$; thus, $L > R > D$ is satisfied.
    
\end{proof}

\begin{lemma}
\label{lemma: two miner Nash}
In the game in Table \ref{table: utility table of two miners bwh betrayal game}, $(B, B)$ is the only pure Nash equilibrium.
\end{lemma}

\begin{proof}

By Lemma \ref{lemma: inequality}, $(C, C)$ is dominated by $(C, B)$ and $(B, C)$. $(B, B)$ dominates $(B, C)$ and $(C, B)$. Therefore, only $(B, B)$ is the Nash equilibrium.

\end{proof}

\noindent For advanced analysis, we assume a general set of BWH miners given as $M = \{m_1, m_2, ..., m_N\}$ where $N>1$.

More generally, we can prove the strategy set that all BWH miners choose the betrayal action is only Nash equilibrium.

\begin{theorem} \label{theorem: n miner Nash}
In the BWH miners game with $N$ miners ($N > 1$), the only Nash equilibrium is for all miners to choose `betray.'
\end{theorem}

\begin{proof}

We can split the set of the BWH miners into two sets based on their actions. Assume there is at least one BWH miner who chooses the action $C$. Then, we denote the set of BWH miners with the action $C$ as $M_1 = \{m_1, m_2, ..., m_k\}$, and the set of BWH miners with the action $B$ as $M_2 = \{m_{k+1}, ..., m_N\}$. Then, for a miner $m_1$, we can change the game to a set of BWH miners $M' = \{m_1, m_{k+1}, ..., m_N\}$. It is the case when $m_1$'s computing power is $p$ and the other miners' computing power is $q$ in Theorem \ref{lemma: two miner Nash}. Therefore, for the miner $m_1$, changing the action to betrayal ($B$) is motivated. Consequently, all the pure strategy sets, including any cooperation ($C$), are not Nash equilibria. 

\end{proof}

\subsection{Pricing on Blocks}

We have already proved that reasonable BWH miners would conduct BDS attacks. Both parties will make a profit or incur no loss due to a result of this trade. Nonetheless, the price of the trade between BDS miners and the victim pool is a range rather than a fixed value. Now we determine an equilibrium price between the BDS miners and the victim pool using a game-theoretic model.

\begin{figure}[tb]
\caption{BDS block pricing game}
\label{fig: block pricing game}
\centering
\includegraphics[width=0.4\linewidth]{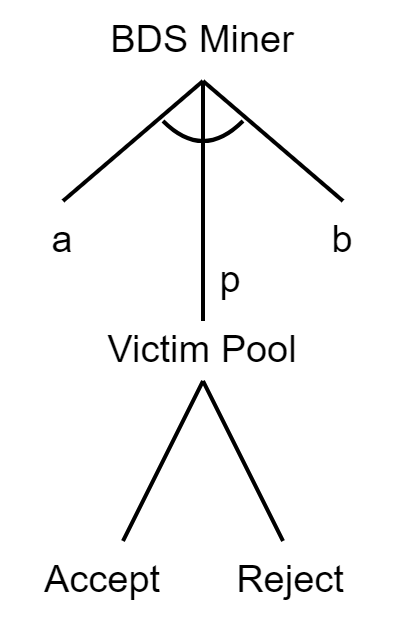}
\end{figure}

Revisiting the assumption for our system model in Section \ref{section: system model and problem}, the victim pool cannot identify or respond to a BWH attack effectively. Therefore, the BDS trade should be proposed by BWH miners within the BWH-attacking pool. If the price of BDS blocks falls within the range specified in Section \ref{section: rationality analysis}, it will always generate more revenue than a BWH attack, which can be regarded as the ultimatum game \cite{guth1982experimental}.

Figure \ref{fig: block pricing game} illustrates the suggested block pricing game. Let $a$ be the lowest boundary, $b$ be the highest boundary in Theorem \ref{theorem: condition statisfaction}, and the proposed price by a BDS miner be $p$. If the victim pool accepts the offer, the BDS miner receives $p$, and the victim pool receives $b - p$. Otherwise, they earn nothing. This ultimatum game can be described as follows:

    \begin{equation}
        f: [a, b] \longrightarrow \{\textsf{Accept} \text{, } \textsf{Reject}  \}
    \end{equation}
    
This game has innumerable Nash equilibria; thus, we refined them using the concept of subgame perfect equilibrium, a refinement framework for dynamic games.

\begin{theorem} \label{theorem: single miner pricing}
In the BDS ultimatum game, (b, \textsf{Accept}) is the only subgame perfect equilibrium.
\end{theorem}

\begin{proof}

Since the range of proposed prices is continuous, it is evident that there is an infinite number of Nash equilibria. The BDS miner proposes $p$, and the victim pool accepts the proposal only if $p$ is included; otherwise, the victim pool rejects the proposal. Nobody can alter their strategies in this situation. Among these Nash equilibria, the strategy where the BDS miner proposes $b$ maximizes his earnings. Therefore, (b, \textsf{Accept}) is the only subgame perfect equilibrium.

\end{proof}

\begin{figure}[tb]
\caption{A Game of BWH Attacker Pool and BWH Miners}
\label{fig: agent game}
\centering
\includegraphics[width=0.9\linewidth]{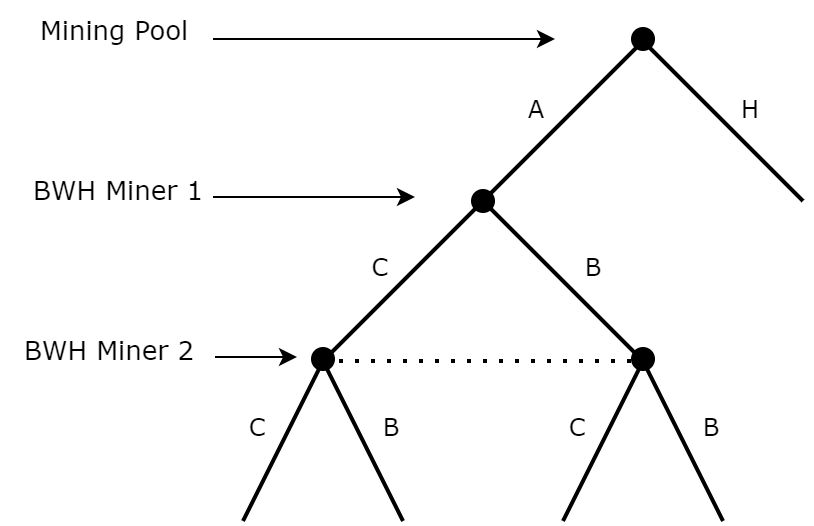}
\end{figure}

\subsection{A Principal-Agent Problem in the BWH Attack}

We analyzed how BWH miners behave in the block-double submission attack situation. Our research indicates that the BWH miners will betray for more profits. If this is the case, the outcome of a BWH attack may differ totally from what the attacker anticipated.

We now construct a new game between a BWH-attacking pool and BWH miners, as illustrated in Figure \ref{fig: agent game}, which is referred to as a BWH-attacking pool and mining agents game. The mining pool may select one of two strategies: the BWH attack (A) or Honest mining (H), with the two sets of BWH miners from the previous game. The BWH attack was designed assuming miners would collaborate with the BWH-attacking pool. The strategy set (A, C, C) with increased profit represents the ideal case (from BWH attacking pool's perspective). The strategy set (A, B, B) is when miners choose the BDS attack. As shown in Section \ref{subsection: cooperation between miners}, the subgame of BWH miners has the only Nash equilibrium to carry out BDS attacks. Therefore, the strategy set (A, C, C) cannot be a Nash equilibrium.

\begin{theorem} \label{theorem: principal-agent game}
The strategy set (H, B, B) is the only subgame Nash equilibrium in the BWH-attacking pool and mining agents game.
\end{theorem}

It implies that mining pools would not want to conduct a BWH attack. The mining pool employs a BWH attack to increase its earnings; however, joining the BWH attack may not be the best decision for BWH miners. It does not work as intended by the BWH-attacking pool. Game theory refers to this type of dilemma as the principal-agent problem. Contrary to the intention of the BWH-attacking pool, a BWH attack will be a strategic failure because reasonable miners seek greater rewards.

%% file: section/6.quantitative.tex
\section{Quantitative Analysis} \label{section: quantitative}

\begin{figure*}[tbp]

  \includegraphics[width=\linewidth]{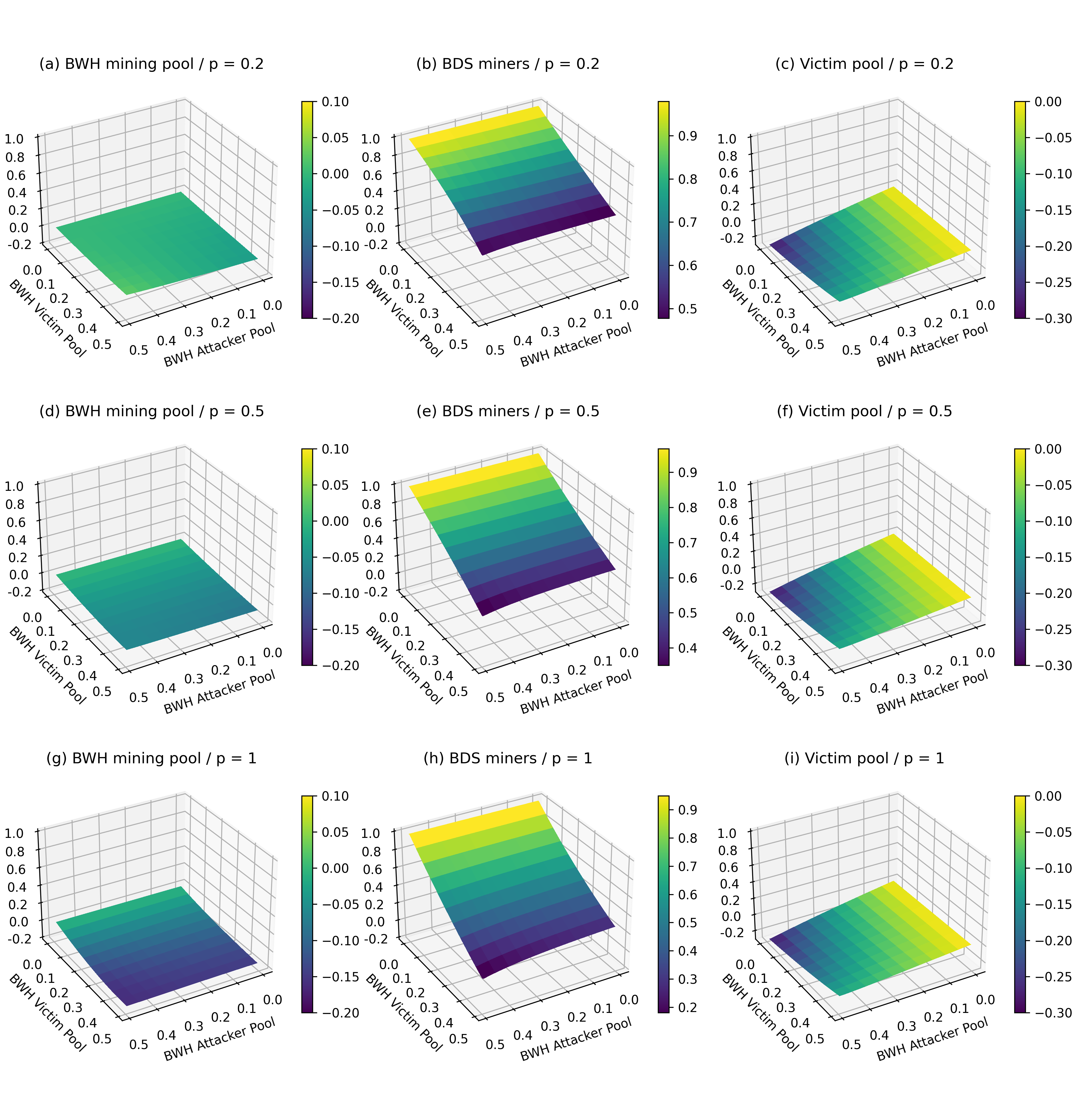}
  \caption{Quantitative analysis results from the BDS with three participation ratios (20\%, 50\%, 100\%): From left to right, each column shows the RERs of the BWH-attacking pool, RERs of the BDS miner, and RERs of the victim mining pool, respectively. It shows that the BDS attack is very profitable. In addition, as the miners betraying the BWH-attacking pool increases, the loss of the BWH attacking pool increases up to nearly 20\%.}
  \label{figure: quantitative result}
  
\end{figure*}

    This section offers a quantitative study of the BDS attack based on different mining power distributions and participation ratios for BDS attacks.

    \begin{equation}
        RER = \frac{R_a - R_h}{R_h}
    \end{equation}

    We provide theoretical results demonstrating the effectiveness of BDS attacks in BWH environments.
    To analyze miners' revenue, we employ the relative extra reward (RER), defined as the ratio of extra revenue to revenue from honest mining. 
    We may derive the following expression, given $R_a$ as the miner's revenue with an attack and $R_h$ as the miner's revenue with honest mining.

    Figure \ref{figure: quantitative result} displays, from left to right, the RERs of the BWH-attacking pool, BDS miners, and victim pool. 20\% of BWH miners participate in the BDS attack in the first row (a) — (c), 50\% in the second row (d) — (f), and 100\% in the third row (g) — (i). First, we can see intuitively that BDS attackers can generate enormous profits. The number of BWH miners who betray the BWH-attacking pool and join the BDS decreases the benefit to be shared; nonetheless, the greatest figure in graph (b) was rounded to the first decimal place, reaching 99.9\% RER. Considering that BWH attacks provide gains of up to 10\% or less, their earnings are astoundingly huge.
    
    Examining the profitability of the BWH-attacking pool reveals that the greater the participation rate in BDS attacks, the lower the effect of BWH, and the lower the RER falls to a negative value. At a BDS participation rate of 100\%, the BWH-attacking pool will experience a loss of up to 20\% and substantial collateral damage. According to our study, involvement in a BDS attack is likely to be a sensible decision from a game-theoretic standpoint, resulting in the BWH attack eventually causing damage to themselves.
    
    Lastly, the RERs of the victim pool have constant fixed values. There are two main factors: First, we have the condition that double-submission of a block does not result in the victim mining pool being lost. Second, according to the analysis in the preceding section, BDS attacks always benefit BDS miners the most from the equilibrium of the ultimatum game; therefore, the victim mining pool neither loses nor gains revenues as a result of existing BWH attacks. Simulations verify the validity of the preceding theoretical analysis in the following section.

%% file: section/7.simulation.tex
\section{Simulation} \label{section: simulation}

This section confirms our analysis of the BDS attack by comparing theoretical expectations with simulated results.

\begin{table}[tb]
\centering
\caption{Approximate Bitcoin power distribution for a month by \url{BTC.com}}
\label{table: Bitcoin mining pool distribution}
\adjustbox{width=\linewidth}{%
\begin{tabular}{C{1cm}C{3.5cm}C{3.5cm}}
\toprule
Rank & Mining Pool         & Computational Power \\ \midrule
1    & Foundry USA  & 18 \%               \\ \midrule
2    & AntPool     & 15 \%               \\ \midrule
3    & F2Pool      & 14 \%               \\ \midrule
4    & Poolin       & 12 \%               \\ \midrule
5    & Binance Pool & 11 \%               \\ \bottomrule
\end{tabular}
}
\end{table}

\begin{figure}[tb]
  \centering
  \includegraphics[width=\linewidth]{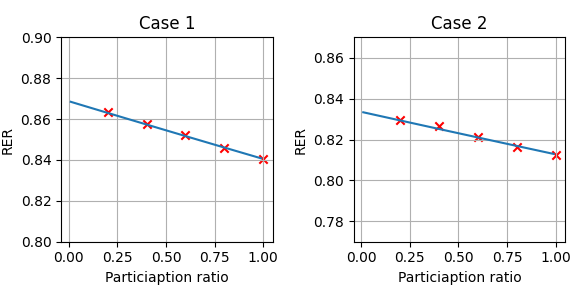}
  \caption{RERs of the BDS miner under a BWH attack. Case 1 and Case 2 were conducted with computational powers (18\%, 15\%) and (12\%, 18\%), respectively. The theoretical values (lines) and simulation results (\textsf{x}-marks) are consistent.}
  \label{figure: simulation}
\end{figure}

\begin{table}[tb]
    \centering
    \caption{The RERs of a BDS miner when the participation ratio is 100\%. Each A (B) number indicates the RERs from the theoretical analysis and simulation, respectively.}
    \label{table: Simulation and Theory Comparison}
    \begin{tabularx}{\linewidth}{>{\centering\arraybackslash}p{1.1cm}>{\centering\arraybackslash}p{1cm}>{\centering\arraybackslash}p{1cm}>{\centering\arraybackslash}p{1cm}>{\centering\arraybackslash}p{1cm}>{\centering\arraybackslash}p{1cm}}
    \toprule
    Case & 20\% & 40\% & 60\% & 80\% & 100\% \\ \midrule
    \# 1  & \begin{tabular}[c]{@{}c@{}}86.36\\ (86.36)\end{tabular} & \begin{tabular}[c]{@{}c@{}}85.80\\ (85.77)\end{tabular} & \begin{tabular}[c]{@{}c@{}}85.23\\ (85.24)\end{tabular} & \begin{tabular}[c]{@{}c@{}}84.67\\ (84.60)\end{tabular} & \begin{tabular}[c]{@{}c@{}}84.11\\ (84.03)\end{tabular}  \\ \midrule
    \# 2  & \begin{tabular}[c]{@{}c@{}}82.98\\ (82.95)\end{tabular} & \begin{tabular}[c]{@{}c@{}}82.56\\ (82.66)\end{tabular} & \begin{tabular}[c]{@{}c@{}}82.14\\ (82.13)\end{tabular} & \begin{tabular}[c]{@{}c@{}}81.73\\ (81.61)\end{tabular} & \begin{tabular}[c]{@{}c@{}}81.32\\ (81.24)\end{tabular}  \\ \bottomrule
    \end{tabularx}
\end{table}

We built a Python Monte Carlo simulator to simulate the double-submission attacks from one pool to another. We employed the actual mining pool environment for simulations. 
Table \ref{table: Bitcoin mining pool distribution} approximates the computing power distribution among the top five mining pools, with which we simulated two BDS attacks.
The first case is when Foundry USA (18\%) attacks AntPool (15\%), and the second case is when Poolin (12\%) attacks Foundry USA (18\%). 

Two simulation results are shown in Figure \ref{figure: simulation}. The blue lines represent theoretical expectations. The RERs fall as the participation ratio increases. The \textsf{x} symbols represent the results from $10^7$ repetitions of the simulation. To validate the effect of the BDS attack, we recreated each scenario with five different BDS participation ratios (20\%, 40\%, 60\%, 80\%, and 100\%). The numbers in Table \ref{table: Simulation and Theory Comparison} provide more precise values up to two decimal places than those shown in Figure \ref{figure: simulation}. 
The simulation results validated the theoretical predictions as depicted in the table and the figures.

%% file: section/8.discussion.tex
\section{Discussion} \label{section: discussion}

This section discusses four aspects of BDS attacks: Adaptability to the FAW attack, viability of the optimal block price, mitigation of BDS attacks, and deception under BDS attacks.

\subsection{Block Double-Submission under Fork After Withholding Attacks}

As introduced in Section \ref{section: related work}, various BWH attack strategies have been studied. Even though this study focuses mainly on the BWH attack, the FAW attack presented by Kwon et al. \cite{kwon2017selfish} is also remarkable.
In the FAW attack, an adversary mining pool infiltrates a victim pool but does not always withhold blocks. If miners who do not belong to the attacking mining pool and the victim mining pool publish a block, the attacking mining pool publishes a delayed fPoW, invoking a fork. If the attacker or the victim wins the fork, the attacker receives additional income. In this way, the FAW attack generates greater revenue than the BWH attack.

The FAW attacks resolved the Prisoners' Dilemma posed by the BWH attacks and provided superior performance. The FAW attacks differentiate from BWH attacks because mining pools with greater mining power profit from the pool vs. pool situation attacks. It indicates that the victim mining pool is disadvantaged even if it detects and counterattacks an FAW attack. Mining pools may then want to attempt the FAW attack, where the rewards of the BDS attack are nearly identical to the one in the BWH case.

Due to the nature of the BDS attack mechanism, if all BWH miners betray, the FAW attack produces the same profits as the BWH attack, which was discussed in the preceding section. Previously, the BDS attack was not intended to submit fPoW to avoid detection by the BWH-attacking pool. In this instance, there is no fPoW for the FAW attacker to attempt forks; hence there is no distinction between this attack and the BWH attack. The BDS attackers can collect the anticipated profits from a BWH attack, while the FAW attackers suffer losses.

\subsection{Viability of Optimal Block Pricing}

Through the subgame Nash equilibrium of the ultimatum game, we determined in Section 5.2 the price at which a BDS attacker sells fPoW to a victim mining pool. Nevertheless, this analysis of the ultimatum game was based on rational decision-making and contained certain inconsistencies with reality. For instance, studies \cite{bolton1995anonymity, guth1982experimental} have revealed that when people are subjected to actual experiments, including the ultimatum games, they prefer to reject offers that provide little value.

Consider three factors that impact ultimatum games in the real world: The first is fairness as a psychological factor \cite{kahneman1986fairness}. For example, suppose that a victim mining pool has already been attacked by a BWH-attacking pool and might have been emotionally insulted. The victim mining pool may reject a BDS proposal if the enhanced advantages disproportional favor the BDS miners.

Another factor is the moral psychology of the proposer. Quantitative evidence indicates that the highest RER for BDS miners is close to 99. Given that the maximum benefit from the BWH attack is close to 10\%, this is a substantial gain. As a result, BDS miners get a substantial profit, and even if they forfeit a portion of that profit, they would be willing to make a few more concessions for the mining pool's collaboration or moral psychology.

The final factor is the magnitude of revenue. According to Forgas and Tan \cite{forgas2013mood}, both the magnitude and ratio of compensation substantially impact on decision-making. This indicates that even if the proportion of benefits acquired by a proposer who received the ultimatum is extremely small, the probability that the proposal will be accepted is large if the benefits themselves are substantial. The size of BDS trades will never be negligible due to Bitcoin's high market capitalization, which continues to rise.

In conclusion, BDS attackers will seek cooperation from the victim mining pool in exchange for concessions rather than the entire theoretical advantages.

\subsection{Detection and Mitigation of Block Double-Submission Attacks}

Multiple detections and evasion measures have been proposed against a BWH attack exploited by BDS attacks, including probabilistic detection, sensing-based detection, and structural prevention techniques.
The BDS attack that exploits trust connections within mining pools might also be detectable or evadable. However, it is impossible to detect attacks using infiltrating sensors when the attacker is a private mining pool. Structural adjustments to the mining protocol are permissible, though not recommended for compatibility. Therefore, probabilistic BDS attack detection is discussed mainly in this study.

Checking the ratio of fPoW to pPoW submissions is a probabilistic technique for identifying BWH attacks. In addition to the probabilistic nature of this technology, the lesser the computational power, the less accurate the detection. Using this point, attackers can make detection difficult by dividing this node's computer power among many miners with low computing power. Similarly, it is possible to establish the mining efficiency probabilistically to detect a BDS attack. Consequently, if a BDS attacker consists of multiple miners as opposed to a single miner, this can diminish the detection accuracy, similar to the detection bypass mechanism for the prior BWH attacks. In addition, the BDS attack can avoid detection by surrendering a small amount of gain, like how BWH attacks are often executed, rather than by operating BDS attacks with optimal efficiency. However, it will still provide considerable additional benefits to the attacks, as the optimal BDS attack nearly doubles the attacker's gain.

\subsection{Deceptive behaviors using Block Double-Submission}

The BWH attack exploits the structural weakness of the mining pool system, whereas the BDS attack exploits the structural weakness of the BWH attack. Therefore, we must investigate the structural flaws of the BDS attack. A major risk exists if a BWH attacker disguises himself and appears to execute a BDS attack against a victim mining pool in an attempt to generate additional revenue.

We can imagine an adversary posing as a BWH attacker and advising a BDS attack to the victim mining pool. The attacking mining pool will gain some profit if the maximum possible benefit of the BDS attack is realized. If this is true, the victim mining pool may be compelled to pay over income even while not under the BWH attack. 
To achieve this, the BDS attacker must prove they are betraying their mining pool.
Because the BDS attacker is a member of the BWH attacker pool, we can establish that the gains of the BDS attack are not shared by bypassing the account of the miner who does not undertake the BWH attack. However, this is not ideal, and it appears that additional research into such cases is required.

%% file: section/10.conclusion.tex
\section{Conclusions} \label{section: conclusion}

BWH attacks are regarded as one of the most dangerous attack techniques in PoW mining. This attack approach undermines sound competition in the PoW blockchain mining ecosystem and can be combined with other attacks. A mining pool is a cooperation between multiple miners, but not all miners in the pool can be trusted. In this work, we presented the BDS attack in which miners betray and trade with a victim pool to earn further benefits within the pool that performs the BWH attack. Notably, we proved that the a trade between BWH miners and a victim pool is viable and have shown theoretically that both parties to the trade could benefit from the BDS attack. In addition, we conducted a game-theoretic analysis of miner behavior and mining pool strategy to estimate the impact of this attack on the BWH attack. The BDS attack is the superior strategy for miners, but the honest mining strategy, instead of the BWH attacks, is the superior strategy for mining pools. In conclusion, we discovered that the BWH attack will not be executed due to a lack of trust between miners and that mining pools will pursue honest mining. Because our attack technique is not impeccable, we believe advanced strategies can be proposed to exploit or defend against it. Therefore, we set the topic aside for our future work.